\PassOptionsToPackage{unicode}{hyperref}
\PassOptionsToPackage{hyphens}{url}
\documentclass[
]{article}
\usepackage[left=2cm, right=2cm]{geometry} 
\usepackage{amsmath,amssymb}
\usepackage{lmodern}
\usepackage{iftex}
\ifPDFTeX
  \usepackage[T1]{fontenc}
  \usepackage[utf8]{inputenc}
  \usepackage{textcomp} % provide euro and other symbols
\else % if luatex or xetex
  \usepackage{unicode-math}
  \defaultfontfeatures{Scale=MatchLowercase}
  \defaultfontfeatures[\rmfamily]{Ligatures=TeX,Scale=1}
\fi
\IfFileExists{upquote.sty}{\usepackage{upquote}}{}
\IfFileExists{microtype.sty}{% use microtype if available
  \usepackage[]{microtype}
  \UseMicrotypeSet[protrusion]{basicmath} % disable protrusion for tt fonts
}{}
\makeatletter
\@ifundefined{KOMAClassName}{% if non-KOMA class
  \IfFileExists{parskip.sty}{%
    \usepackage{parskip}
  }{% else
    \setlength{\parindent}{0pt}
    \setlength{\parskip}{6pt plus 2pt minus 1pt}}
}{% if KOMA class
  \KOMAoptions{parskip=half}}
\makeatother
\usepackage{xcolor}
\usepackage{graphicx}
\makeatletter
\def\maxwidth{\ifdim\Gin@nat@width>\linewidth\linewidth\else\Gin@nat@width\fi}
\def\maxheight{\ifdim\Gin@nat@height>\textheight\textheight\else\Gin@nat@height\fi}
\makeatother
\setkeys{Gin}{width=\maxwidth,height=\maxheight,keepaspectratio}
\makeatletter
\def\fps@figure{htbp}
\makeatother
\ifLuaTeX
  \usepackage{selnolig}  % disable illegal ligatures
\fi
\IfFileExists{bookmark.sty}{\usepackage{bookmark}}{\usepackage{hyperref}}
\IfFileExists{xurl.sty}{\usepackage{xurl}}{} % add URL line breaks if available
\hypersetup{
  hidelinks,
  pdfcreator={LaTeX via pandoc}}

\author{}
\date{}

\begin{document}

\textbf{Phase transformation kinetics in MoS\textsubscript{2} governed
by S--S repulsive interactions and defect--interface compatibility}

Pai Li\textsuperscript{1}, Ziao Tian\textsuperscript{1}, ZengFeng
Di\textsuperscript{1}*, Feng Ding\textsuperscript{2}*

\emph{\textsuperscript{1}State Key Laboratory of Materials for
Integrated Circuits, Shanghai Institute of Microsystem and Information
Technology, Chinese Academy of Sciences, Shanghai 200050, China}

\emph{\textsuperscript{2}Suzhou Laboratory, Suzhou, 215123, China}

*Corresponding Author

Email: zfdi@mail.sim.ac.cn; dingf@szlab.ac.cn

\textbf{Abstract}

The metastable T' phase in monolayer MoS$_2$ exhibits remarkable
persistence despite a strong thermodynamic driving force toward the
stable H phase. Using machine learning--accelerated molecular dynamics
and first-principles calculations, we reveal that this kinetic arrest
originates from repulsive S--S interactions, which impose high energy
barriers during both nucleation and grain boundary propagation. While
sulfur vacancies can alleviate these barriers in certain interfaces,
they fail to accelerate transformation at the most stable
interface, ZZ-Mo\textbar-, due to their thermodynamic instability
there. Instead, vacancies migrate into the T' phase, leaving the
advancing front defect-free. Direct simulations of nanostructures
confirm that H-phase nucleation initiates at corners or edges, and all
observed growth fronts adopt the ZZ-Mo\textbar- configuration,
consistent with its low interfacial energy but slow kinetics. Our work
establishes that phase transformation in 2D materials is governed not by
global defect concentration, but by the local compatibility between
defects and moving interfaces, offering a new paradigm for controlling
structural transitions through interface-specific design.

\textbf{Introduction}

The controlled phase transformation between semiconducting and metallic
polymorphs is a cornerstone for engineering functional two-dimensional
(2D) materials.\textsuperscript{1} In monolayer MoS\textsubscript{2},
the thermodynamically stable H phase coexists with a metastable metallic
T' phase, which exhibits promising properties for catalysis,
electronics, and topological devices.\textsuperscript{2--8} Despite a
substantial energy difference due to electronic configurations (0.55
eV/f.u. for the T' phase and 0.85 eV/f.u. for the ideal T phase with
octahedra O\textsubscript{h} symmetric units),\textsuperscript{9,10} the
T' phase can be obtained from the H phase by temporarily reversing the
thermodynamic stability order via lithium
intercalation,\textsuperscript{11} electron beam
irradiation,\textsuperscript{12} argon plasma
treatment,\textsuperscript{13} or anion extraction.\textsuperscript{14}
It can also be synthesized directly through chemical vapor
deposition\textsuperscript{4} or wet-chemical
routes,\textsuperscript{15} with the T' phase stabilized by the
substrate.\textsuperscript{16} Meanwhile, the pure metastable T' phase
MoS\textsubscript{2} can persist for months at room
temperature\textsuperscript{3,17}---a kinetic stability that defies
simple expectations given the seemingly modest atomic rearrangement
required for transformation.

Recent theoretical studies have identified various grain boundary
structures and highlighted differences in their
mobility.\textsuperscript{18--20} Zhao et al. revealed that zigzag-type
boundaries should dominant in steady states.\textsuperscript{19} Zhou et
al. further revealed that among zigzag-type boundaries, the S-orientated
interfaces are more immobile due to the high formation energies of kinks
and thus dominate the physical properties of the whole
heterostructures.\textsuperscript{20} It was reported that S vacancy
facilitate the transformation from H to T\textquotesingle{} phase
theoretically and experimentally.\textsuperscript{13,14,21} However, as
a typical kinetics-dominated process, a unified mechanism linking
atomic-scale interactions, defect thermodynamics, and macroscopic
transformation kinetics remains elusive. In particular, whether---and
how---sulfur vacancies assist the T' → H transformation has never been
critically examined from the perspective of defect stability at moving
interfaces.

Here, we combine density functional theory (DFT), crystal orbital
Hamilton population (COHP) analysis, and machine learning--accelerated
molecular dynamics to unravel the origin of kinetic arrest in monolayer
MoS\textsubscript{2}. We demonstrate that repulsive S--S interactions impose universal
energy barriers during both nucleation and interface propagation. More
strikingly, we reveal that sulfur vacancies cannot promote
transformation at the most stable and prevalent interface
(ZZ-Mo\textbar-) because they are thermodynamically unstable there and
rapidly migrate into the T' phase. Direct simulations of nanostructures
confirm that H-phase growth exclusively proceeds via ZZ-Mo\textbar-
boundaries, yet remains slow due to the absence of persistent vacancies.
Our work establishes that the efficacy of defects in driving phase
transformations is governed not by their global concentration, but by
their local compatibility with the advancing interface---a principle
with broad implications for defect engineering in 2D materials.

\textbf{Results}

\textbf{S-S Repulsive Interaction at T'/H Grain Boundaries}

To systematically investigate the atomic structure and bonding nature of
the T/H interfaces, we constructed seven representative coherent grain
boundaries (CGBs) between the H and T phases, including both zigzag- and
armchair-oriented variants: ZZ-S\textbar-, ZZ-Mo\textbar+,
ZZ-S\textbar+, ZZ-Mo\textbar-, AC\textbar0, AC\textbar-, and
AC\textbar+. These configurations are based on the crystallographic
alignment proposed in prior work,\textsuperscript{19} with slight
modifications to the AC interface naming. Here, ``+(-)'' in each
notation indicates the interface is S rich (depletion). The full set of
initial and optimized atomic structures is provided in Supplementary
Information Fig. S1.

In the pristine (unrelaxed) and S rich interface models, the interfacial
S--S distance is drastically reduced to approximately 1.84 Å, far
shorter than the equilibrium S--S separation in bulk H- or T-phase MoS\textsubscript{2}
(\textasciitilde3.2 Å) and close to the S-S covalent bond lengths (1.89
Å) in S\textsubscript{2} molecule. Upon full structural relaxation, the
system responds by pushing the S atoms apart: the S--S distances
increase to 2.68 Å in ZZ-Mo\textbar+, 2.79 Å in ZZ-S\textbar+, and 2.57
Å in AC\textbar+ (Fig. 1), much longer than a normal S-S bond. Despite
this relief, these post-relaxation distances remain significantly
shorter than the van der Waals contact ($\sim$3.6 Å) and well below the
intrinsic S--S spacing in either pristine H or T phase, indicating
persistent strain and repulsion at the interface. This structural
response manifests as pronounced local bending or buckling of the
monolayer, particularly evident in the side views of the relaxed
interfaces (Fig. 1). The deformation reflects the energetic cost of
accommodating the misaligned sulfur sublattices---a cost primarily
driven by S--S repulsion.

\begin{center}
    \includegraphics[width=4in,height=4in]{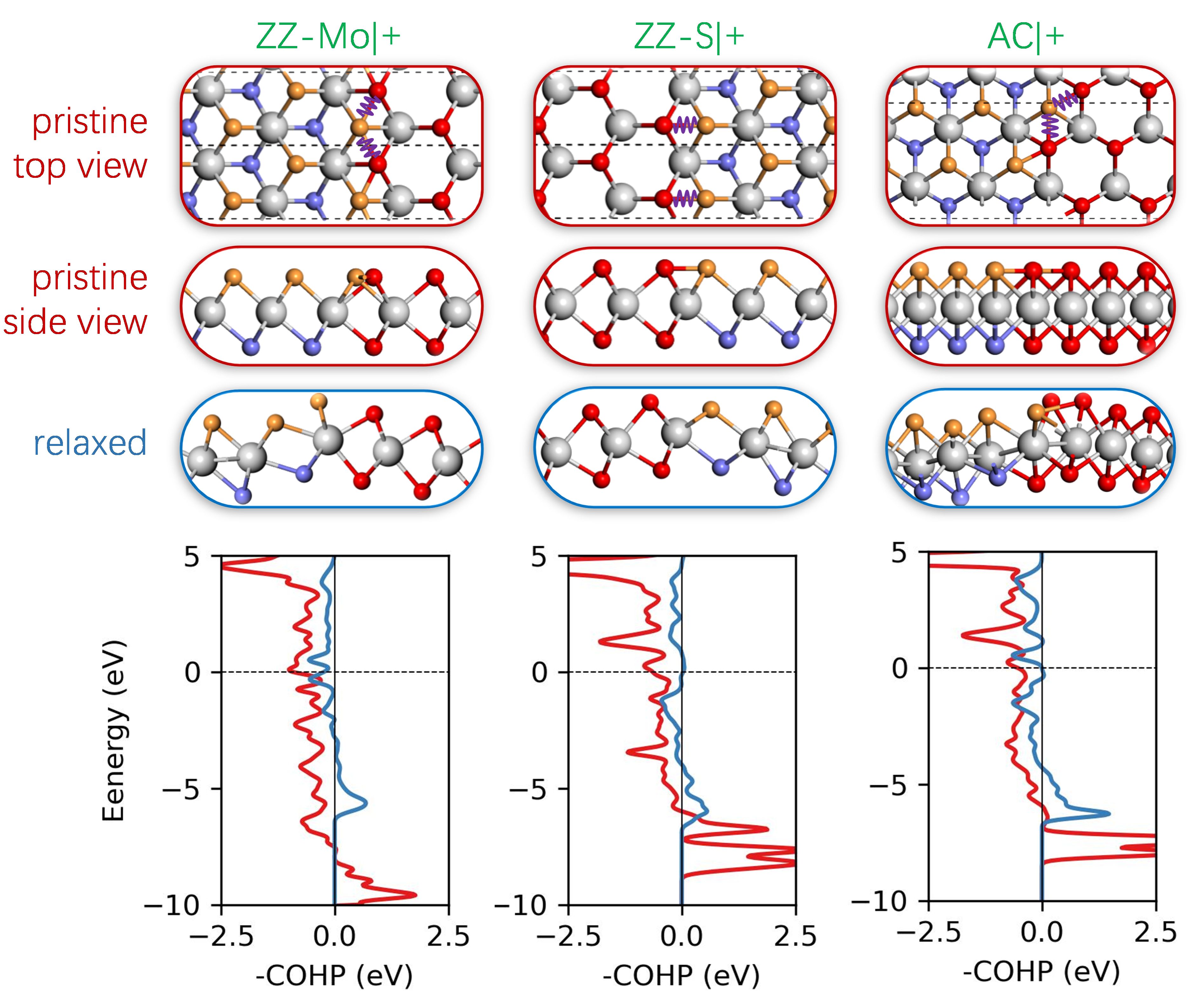}
\end{center}

\textbf{Fig. 1 \textbar{} DFT analysis of S--S repulsive interactions at
H--T' grain boundaries.} Top and side views of the atomic structures
before relaxation (red frames), and side views of the fully optimized
geometries after relaxation (blue frames) for S-rich interfaces:
ZZ-Mo\textbar+, ZZ-S\textbar+, and AC\textbar+. Repulsive S--S
interactions are indicated by wavy lines between closely spaced S atoms.
Mo atoms are gray; upper and lower S atoms in T-phase area are orange
and blue, respectively; overlapping S atoms (in top view) in H-phase
area are colored red. Below each atomic model are the corresponding
crystal orbital Hamilton population (COHP) curves for S--S pairs,
plotted before (red) and after (blue) relaxation. Negative values
indicate antibonding character, which correlates with repulsive S--S
interactions.

To quantify this interaction, we performed crystal orbital Hamilton
population (COHP) analysis\textsuperscript{22} on S-rich interfaces
(ZZ-Mo\textbar+, ZZ-S\textbar+, AC\textbar+) before and after
relaxation. As illustrated in Fig. 1, the COHP curves show a pronounced
antibonding character across a wide energy range below the Fermi level
in the unrelaxed configurations (red lines). After relaxation (blue
lines), the S--S distance increases and the magnitude of the repulsive
contribution decreases, confirming that the system avoids the
high-energy configuration through atomic displacement.

Importantly, the presence of such S--S repulsion is not limited to
specific interface types but appears universally across all point
defects or line defects due to the ionic nature of chemical bonds in
MoS\textsubscript{2}. These findings indicate that the formation of
unfavorable S--S neighbors may act as a kinetic barrier, impeding both
nucleation and growth of the H phase within the T' matrix.

\textbf{Machine learning force field (MLFF) training for phase
transition simulations}

To enable large-scale atomistic simulations of the phase transformation
in monolayer MoS\textsubscript{2} over larger spatial and time scales, we developed a
high-accuracy MLFF based on deep neural networks using the DeepMD-kit
framework.\textsuperscript{23} The MLFF was trained to reproduce the
atomic forces and energies obtained from DFT calculations across a broad
configuration space (Fig 2a,b) that includes not only pristine phases
but also defective, strained, and interfacial structures critical for
phase transition dynamics (more details can be found in Supplementary
Information). The performance of the MLFF is quantified by root mean
square errors (RMSE): 6.53 meV/atom for energy and 146 meV/Å for
forces---indicating excellent agreement with DFT. The MLFF is also
validated by the phonon spectra and thermodynamics stability
calculations (Fig 2c,d).

\includegraphics{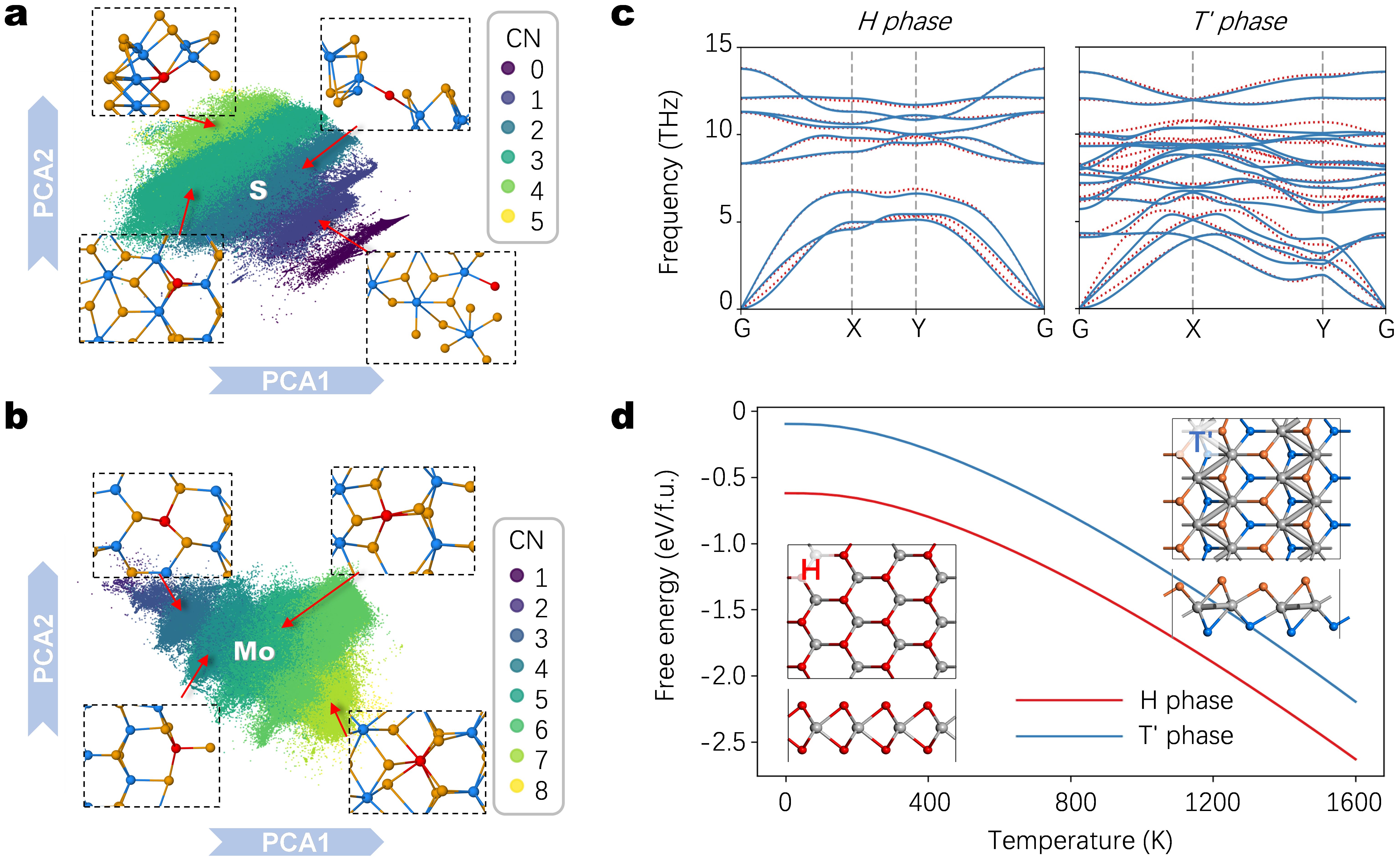}

\textbf{Fig. 2 \textbar{} Training and validation of the MLFF and phase
stability of MoS\textsubscript{2}.} (a, b) Principal component analysis
(PCA) of local atomic environments in the training set, colored by Mo--S
coordination number (CN): (a) for S-centered and (b) for Mo-centered
environments. Insets show representative local structures; S atoms are
orange, Mo atoms are blue, and the central atom (S in a, Mo in b) is
highlighted in red. (c) Phonon dispersion of H- and T'-phase MoS\textsubscript{2}
computed using the MLFF (blue solid lines) and DFT (red dashed lines).
(d) Helmholtz free energy as a function of temperature for the H and T'
phases. The insets show supercell structures of the H phase and the
distorted T' phase, with Mo--Mo dimerization in the T' phase emphasized
by bold bonds.

\textbf{S-S interaction as the bottleneck in H-phase nucleation within
T}\textquotesingle{} \textbf{matrix}

To elucidate the nucleation mechanism, we performed structural
optimization using our trained MLFF, focusing on the emergence of a
compact H-phase nucleus in a T matrix with minimal number of S-S pairs.
As shown in Fig. 3a, the nucleation process proceeds through a sequence
of sulfur atom hops, where each step leads to the formation of a new
hexagonal ring characteristic of the H phase. The progression is marked
by increasing nucleus size, defined by the number of such hexagonal
units.

\includegraphics{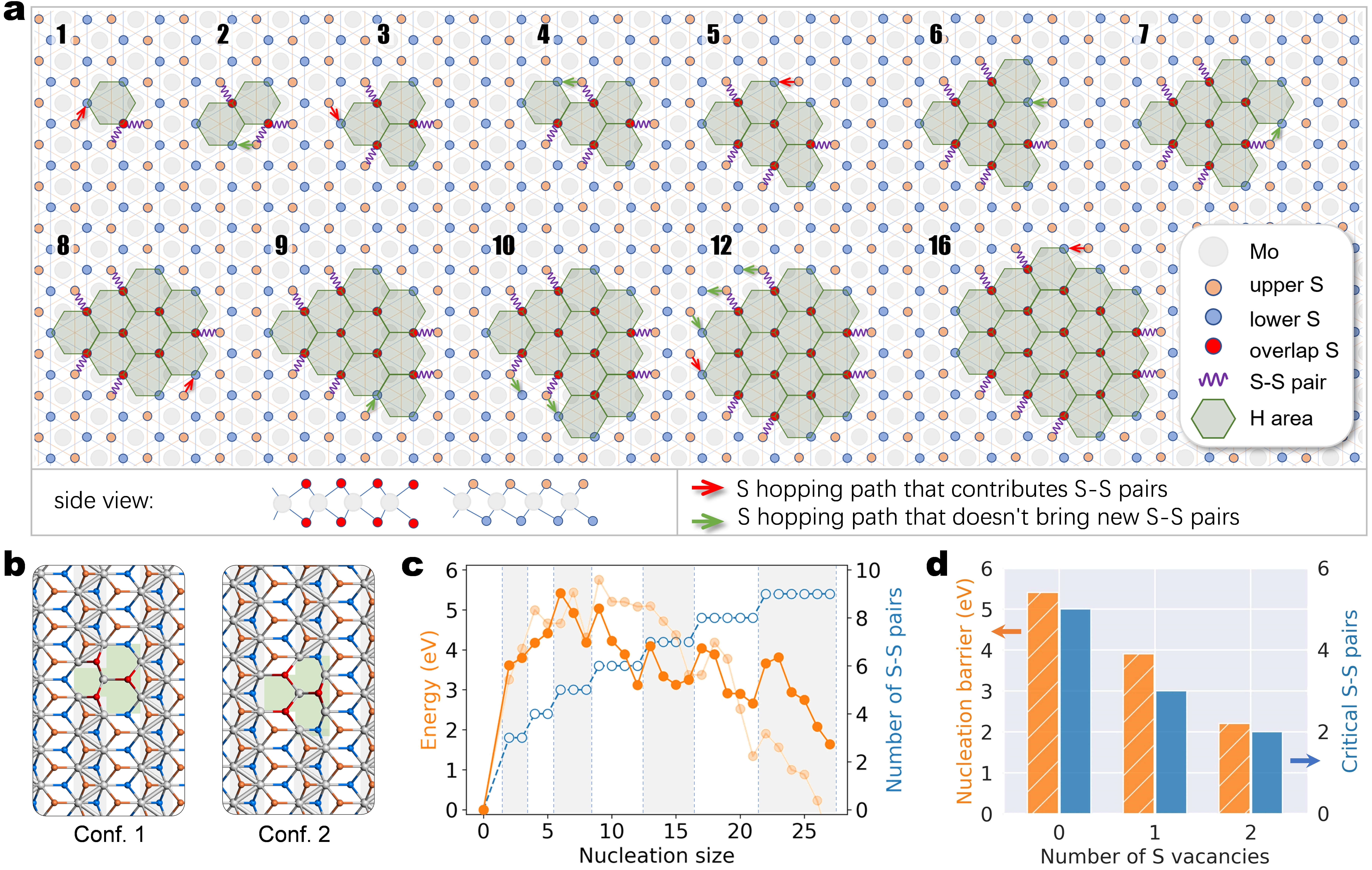}

\textbf{Fig. 3 \textbar{} Nucleation of the H phase in T-phase
MoS\textsubscript{2}.} (a) Schematic illustration of homogeneous
nucleation within the T matrix. Each nucleus is labeled by the number of
H-like hexagonal rings it contains. Red (green) arrows indicate S atom
hops that create (avoid) new S--S pairs; wavy lines denote repulsive
S--S interactions. Upper and lower S atoms are shown in orange and blue
in T-like areas, respectively; overlapping sites (in top view) are
colored red in H-like areas. (b) Two distinct local configurations for
size-3 nuclei in the T' phase, arising from its reduced symmetry
compared to the ideal T structure. This structural degeneracy persists
across all nucleus sizes. (c) Energy profiles (orange and light orange)
for the two nucleation pathways in T' due to reduced symmetry,
corresponding to different initial configurations, overlaid with the
number of S--S pairs (blue curve). The energy increases when new S--S
pairs form and decreases when the nucleus grows without introducing
additional repulsive contacts. (d) Dependence of the nucleation barrier
(left axis) and the number of S--S pairs at the critical nucleus size
(right axis) on sulfur vacancy (V\textsubscript{S}) concentration. The critical S--S
pair count refers to the number present in the nucleus at the transition
state.

A striking feature observed across all nucleation pathways is the strong
correlation between the number of S--S pairs and the system energy. As
illustrated in Fig. 3c, when the nucleus grows without forming new S--S
pairs (indicated by green arrows), the total energy decreases steadily.
However, at certain critical steps---marked by red arrows---the hopping
of a sulfur atom creates a new S--S pair, leading to a sharp increase in
energy. This pattern repeats periodically: energy rises with S--S
creation, then drops again upon further growth without additional
repulsive interactions.

This behavior is consistent with the electronic origin of the S--S
repulsion identified earlier: close proximity of sulfur atoms results in
strong antibonding interactions, which are energetically costly. While
the number of S-S pairs are inevitably increase with the nucleation
process, each S--S pair formed during nucleation acts as a local energy
barrier, making the overall process highly sensitive to the sequence of
atomic movements.

Interestingly, the presence of sulfur vacancies significantly alters
this landscape. In systems with one or two S vacancies (Supplementary
Information Figs. S3--S4), the number of S--S pairs required for
nucleation is reduced. For instance, in the case of two S vacancies, the
same-sized nucleus contains fewer S--S contacts, resulting in a lower
energy profile (Fig. 3d). More importantly, the nucleation barrier is
dramatically reduced: from \textasciitilde5.4 eV in pristine T' to
\textasciitilde2.1 eV with two vacancies (Fig. 3d). This reduction is
directly linked to the suppression of S--S pairing due to missing sulfur
atoms, which effectively ``opens up'' space for favorable atomic
reorganization.

We also identify two distinct local configurations for each nucleus size
in the T' phase (Fig. 3b), reflecting the lower symmetry of T' compared
to ideal T phase. These configurations differ slightly in their
coordination environments and thus exhibit subtle differences in energy
and S--S contact counts. Nevertheless, both follow the same general
trend: energy spikes coincide with S--S pair formation. Therefore, the
S--S repulsive interaction is not just a passive consequence of atomic
packing---it is the dominant factor controlling the kinetics of phase
transformation.

\textbf{S-S pairs govern grain boundary propagation kinetics}

The transformation from the metastable T' phase to the thermodynamically
stable H phase in monolayer MoS\textsubscript{2} is not only hindered at the nucleation
stage but also proceeds slowly during the propagation of the H/T
interface. To investigate this process, we examined the interface energy
(Fig. 4a) of different coherent grain boundaries (CGBs) and calculated
the energy barriers of interface propagation for the four
representatives with lowest interface energies in the reasonable S
chemical potential range: ZZ-Mo\textbar-, AC\textbar-, ZZ-S\textbar+,
and ZZ-S\textbar-. As illustrated in Fig. 4b-e, each propagation step
involves a sequence of sulfur atom hops that either create or avoid new
S--S pairs. The energy profiles below each configuration show a clear
pattern: when a hop leads to the formation of an S--S pair (indicated by
red arrows in Fig. 4), the system experiences a large barrier and a
significant reaction energy. Specifically, the energy barrier of kink
nucleation is up to 1.08 eV in the case of ZZ-Mo\textbar- due to the
strong S-S pair repulsive interactions. In contrast, hops that do not
generate new S--S pairs (green arrows) proceed with minimal energy cost,
often accompanied by energy release due to favorable bonding
reorganization. For example, in the AC\textbar- interface (Fig. 4c), the
first hop introduces a small energy barrier of 0.014 eV, followed by a
large exothermic drop of 0.43 eV. Similarly, in ZZ-S\textbar+ (Fig. 4d),
although the first hop generates a high barrier of 0.95 eV due to S--S
pair creation, the following steps reduce the energy progressively
without new pair creation as kinks advance. The situation of
ZZ-S\textbar- interface is similar with AC\textbar-, where no S-S pair
creation in the whole interface propagation process, and the barrier is
relatively low.

\includegraphics{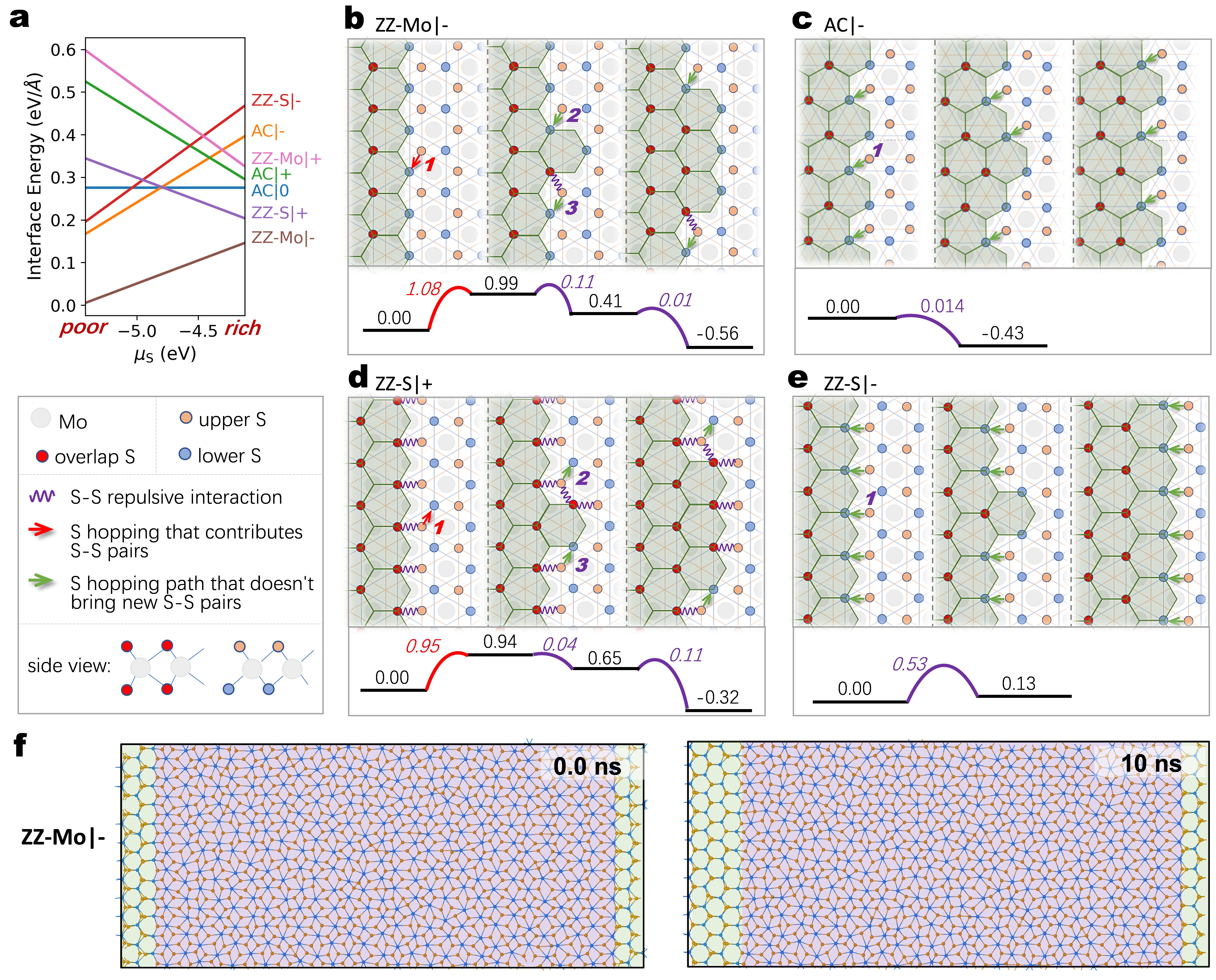}

\textbf{Fig. 4 \textbar{} Analysis of H--T' interface propagation.} (a)
Interface energy of various H--T' boundaries as a function of sulfur
chemical potential ($\mu$\textsubscript{S}), showing the thermodynamic stability of
different interfaces under S-poor (with Mo metal as energy reference)
and S-rich (with S\textsubscript{8} molecule as energy reference)
conditions. (b--e) Atomic-scale mechanisms of interface motion for (b)
ZZ-Mo\textbar-, (c) AC\textbar-, (d) ZZ-S\textbar+, and (e)
ZZ-S\textbar-. The red (green) arrows denote S atom hops that create
(avoid) new S--S pairs; wavy lines highlight repulsive S--S pairs. Upper
and lower S atoms in T' areas are shown in orange and blue,
respectively; overlapping (in top view) S atoms in H areas are colored
red. Energy barriers (in eV) along each pathway are indicated below the
schematics. (f) Molecular dynamics snapshots of ZZ-Mo\textbar- interface
propagation at 1200\,K. Left: initial configuration after equilibration;
right: configuration at 10\,ns. The left interface is mobile
(ZZ-Mo\textbar-), while the right one (ZZ-S\textbar-) is fixed by
constraining the in-plane positions of neighboring S atoms.

Beyond the kinetic control by S--S repulsion, the lower symmetry of the
T' phase compared to the ideal T phase introduces additional complexity
in GB structures. This leads to multiple possible configurations for the
same interface type, some of which may be locally metastable considering
the complex Mo-Mo interaction.\textsuperscript{24} However, these
structural variations do not fundamentally alter the role of S--S pairs
in governing propagation kinetics---they remain the dominant factor
determining energy barriers. The previous results are based on the most
stable interface configurations. More discussions on these structural
variations can be found in Supplementary Information part.

We also performed MD simulations to confirm these results. As shown in
Fig.4f, the propagation of ZZ-Mo\textbar- is so slow that it only
advanced a single lattice unit in 10 ns at 1200 K in our simulations.
ZZ-S\textbar+ has also a very low mobility, while other interfaces show
very fast propagation (Supplementary Information Fig. S9).

These results demonstrate that the dynamics of phase boundary motion are
not solely determined by thermodynamic driving forces but are critically
regulated by local electronic interactions. Specifically, the avoidance
of S--S contact becomes a selection criterion for fast-propagating
interfaces. Thus, controlling sulfur availability or engineering defect
structures that minimize S--S pairing could provide a route to
accelerate phase transformation in 2D materials.

\textbf{Sulfur vacancies cannot facilitate transformation by alleviating
S-S repulsion}

While defects are widely believed to act as catalysts for phase
transformations in materials by reducing local strain or providing
migration pathways,\textsuperscript{25,26} our findings reveal a
counterintuitive truth: the presence of sulfur vacancies does not
universally accelerate the grain boundary propagation in
MoS\textsubscript{2} in the T' → H transformation. This conclusion
arises from a detailed analysis of two representative coherent grain
boundaries (CGBs): ZZ-S\textbar+ and ZZ-Mo\textbar-. Both interfaces
exhibit distinct behaviors under identical sulfur vacancy conditions,
despite both being capable of forming S--S pairs during propagation.

At first glance, it appears that the propagation barriers for both
ZZ-Mo\textbar- and ZZ-S\textbar+ are lowered by introducing a sulfur
vacancy (Fig. 5a,b). For ZZ-Mo\textbar-, the vacancy reduces the kink
generation barrier from 1.08 eV to 0.53 eV. However, ZZ-Mo\textbar- is
an S-depleted interface, and a sulfur vacancy may not be stable
there---unlike at the S-rich ZZ-S\textbar+ interface. To assess this, we
calculated the formation energy of a sulfur vacancy (V\textsubscript{S})
at different locations. As shown in Fig. 5c, the vacancy formation
energy is highest in the H phase, followed by the ZZ-Mo\textbar-
interface, then the T' phase, and lowest at the ZZ-S\textbar+ interface.
This trend implies that V\textsubscript{S} can readily ''ride'' along
the ZZ-S\textbar+ interface but is energetically unfavorable at
ZZ-Mo\textbar-. Crucially, the formation energy of a sulfur vacancy at
the ZZ-Mo\textbar- interface (2.03 eV, referenced to the S\textsubscript{8} molecule) is
significantly higher than that in the T' phase (1.28 eV). Consequently,
the equilibrium vacancy concentration at ZZ-Mo\textbar- is estimated to
be $\sim$2.4 $\times$ 10\textsuperscript{-13} times lower than in the T' phase at room temperature.
Therefore, even under conditions of high sulfur vacancy concentration,
the ZZ-Mo\textbar- interface remains essentially vacancy-free, rendering
sulfur vacancies ineffective in promoting its propagation.

\begin{center}
    \includegraphics[width=4in,height=4in]{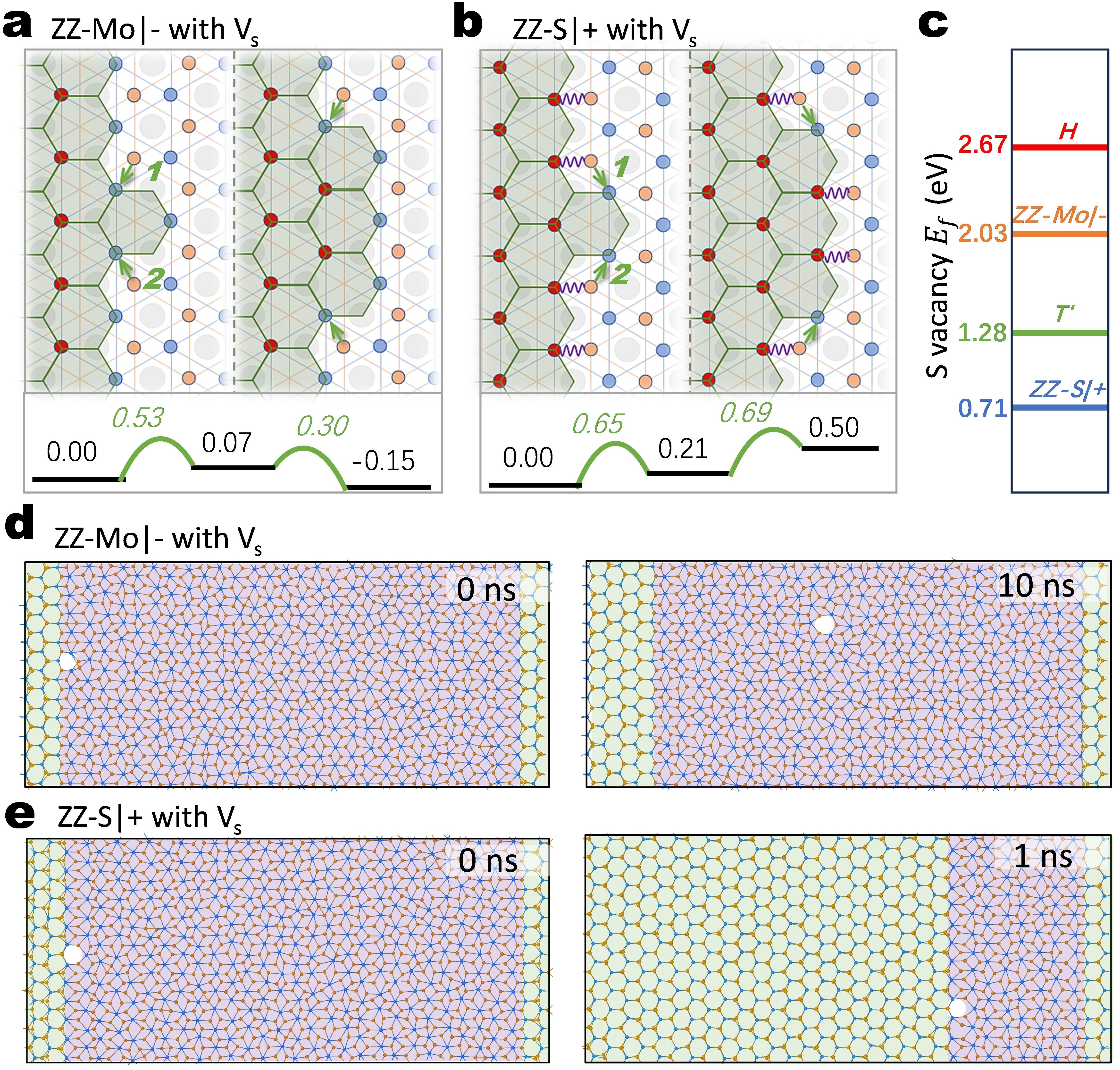}
\end{center}

\textbf{Fig. 5 \textbar{} H--T' interface propagation facilitated by
sulfur vacancies (V\textsubscript{S}).} (a) Propagation mechanism of the ZZ-Mo\textbar-
interface with one V\textsubscript{S}. Elementary steps are illustrated in the inset,
with reaction energies and energy barriers shown below. (b) Propagation
mechanism of the ZZ-S\textbar+ interface with one V\textsubscript{S}. (c) Formation
energy of V\textsubscript{S} at different positions in the MoS\textsubscript{2} lattice.
(d,e) Molecular dynamics simulations at 1200\,K for ZZ-Mo\textbar- and
ZZ-S\textbar+ interfaces with V\textsubscript{S}. The H and T' phases are colored green
and purple, respectively; sulfur vacancies are highlighted in white.
While ZZ-Mo\textbar- and ZZ-S\textbar+ are the most stable interfaces
with low intrinsic mobility, the presence of a V\textsubscript{S} significantly enhances
the mobility of ZZ-S\textbar+ by more than an order of magnitude
compared to the vacancy-free case. In contrast, the V\textsubscript{S} in ZZ-Mo\textbar-
readily diffuses away from the interface, leaving the interface
immobile.

Our MD simulations confirm this. In the presence of V\textsubscript{S},
the ZZ-S\textbar+ interface rapidly advances within 1 ns (Fig. 5d) which
is much faster than the propagation of the perfect ZZ-S\textbar+
interface (Supplementary Information Fig. S9). Meanwhile, we note the
vacancy follows the interface propagation. In contrast, for
ZZ-Mo\textbar-, when a sulfur vacancy is initially placed at the
ZZ-Mo\textbar- interface, it quickly diffuses into the T domain and the
interface propagation is still very difficult. Thus, while
V\textsubscript{S} can alleviate S--S repulsion in certain interfaces
like ZZ-S\textbar+, they fail to do so in the most stable interface
(ZZ-Mo\textbar-) due to instability, demonstrating that defect-mediated
acceleration is not guaranteed---it requires both energetic favorability
and spatial persistence.

\textbf{Direct Observation of Spontaneous H-Phase Nucleation and Growth}

To validate our mechanistic understanding of the T → H phase
transformation, we performed large-scale molecular dynamics (MD)
simulations at 1200 K on several representative MoS\textsubscript{2} nanostructures: a
triangular island with zigzag edges, an armchair-edge ribbon, and a
zigzag-edge ribbon. These systems were initialized in the metastable T
phase and allowed to evolve freely, enabling us to directly observe the
spontaneous nucleation and growth of the H phase under realistic thermal
conditions.

As shown in Fig. 6a, in the triangular island with zigzag edges, the
first H-phase nucleus appears at one of the corners, consistent with the
less coordination and higher activity. Once nucleated, the H domain
grows outward along the edges, where kinks are generated and propagate.
Importantly, all observed grain boundaries between the H and T phases
are identified as ZZ-Mo\textbar-, matching our earlier prediction that
this interface is both thermodynamically stable and kinetically
accessible under these conditions. The schematic inset illustrates the
two key stages: (i) kink generation at the edge, and (ii) stepwise
propagation of the ZZ-Mo\textbar- interface, which proceeds slowly but
steadily due to the energy barriers associated with S--S pair formation.

\includegraphics{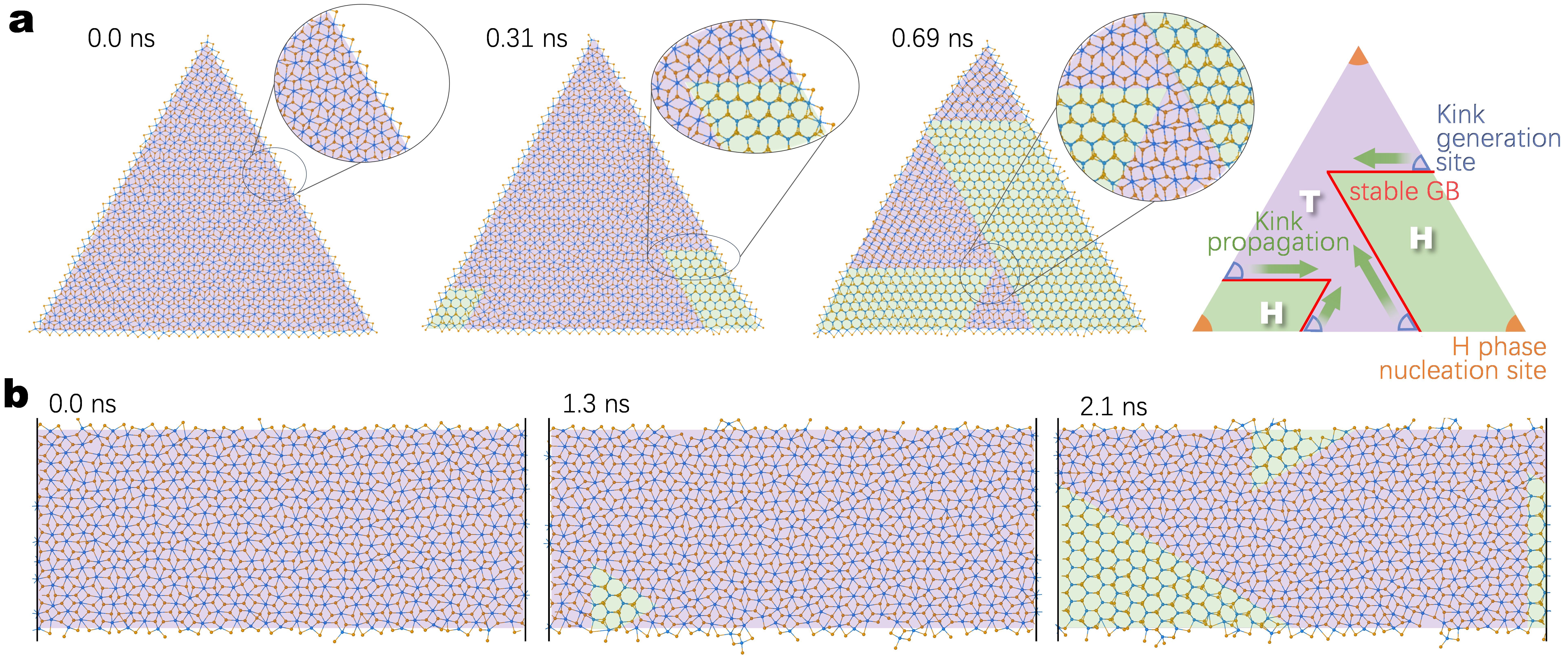}

\textbf{Fig. 6 \textbar{} Snapshots of MoS\textsubscript{2} phase
transformation from T' to H phase at 1200\,K.} (a) MD simulation of a
triangular MoS\textsubscript{2} island with zigzag edges. The rightmost inset
illustrates the two key stages of interface evolution: kink generation
at the stable grain boundary (GB) and subsequent kink migration into the
T' phase. (b) Simulation of a MoS\textsubscript{2} ribbon with armchair
edges. Time stamps above each snapshot indicate the simulation time;
time zero corresponds to the completion of equilibration at 1200\,K,
marking the start of the production run. The T' and H phases are colored
purple and green, respectively.

Similarly, in the armchair-edge ribbon (Fig. 6b), the H phase nucleates
preferentially at the edges and propagates inward. Again, the resulting
interface is exclusively ZZ-Mo\textbar-, confirming its dominance across
different geometries. Notably, no other interfaces such as ZZ-S\textbar+
or AC\textbar± are observed during the entire simulation time,
suggesting that they are either energetically less favorable or
kinetically inaccessible (kinetic Wulff
construction\textsuperscript{27,28}) under these conditions. In
contrast, the zigzag-edge ribbon shows no sign of phase transformation
even after 10 ns at 1200 K. This observation is consistent with our
analysis: while the ZZ edge provides a favorable site for kink
formation, the absence of sufficient sulfur vacancies or favorable
nucleation pathways prevents the initiation of H-phase growth.
Furthermore, the lack of corner sites may reduce the probability of
spontaneous nucleation.

\textbf{Discussion}

The phase transformation from metastable T' to stable H-phase monolayer
MoS\textsubscript{2} proceeds via nucleation followed by interface propagation. In
pristine systems, nucleation has a high barrier of 5.4\,eV, reduced to
3.9\,eV with one V\textsubscript{S} and 2.2\,eV with two. However,
V\textsubscript{S} clustering in T' phase is thermodynamically
disfavored:\textsuperscript{29} due to the ionic nature of Mo--S
bonding, V\textsubscript{S} carries an effective positive charge, making
pairwise proximity energetically costly (+0.37\,eV; Supplementary
Information Fig. S10). Thus, the practical effect of V\textsubscript{S}
on nucleation is limited.

After nucleation, propagation occurs via kink generation and migration
along the grain boundary. The ZZ-Mo\textbar- interface is both the most
thermodynamically stable and the dominant configuration in simulations,
which is also consistent with experiments.\textsuperscript{30} But it is
S-depleted and cannot stably host V\textsubscript{S}, which instead
migrates into the T' phase. Kink generation on ZZ-Mo\textbar- requires
forming an unfavorable S--S pair, yielding a high barrier of 1.08\,eV,
whereas kink migration (involving no new S--S contacts) proceeds with
only 0.20\,eV. Since kink generation can occur anywhere along the
interface, its rate scales with interface length, while full kink
traversal time decreases with length. Hence, under normal conditions,
propagation is limited by the slower kink generation step (see Methods
for more discussion).

These results establish a general principle: a defect promotes
transformation only if it is thermodynamically stable at the moving
interface. While many systems meet this criterion, MoS\textsubscript{2} is an
exception---the very stability of ZZ-Mo\textbar- renders it inert to
vacancy assistance. Consequently, despite a strong driving force
(0.55\,eV/f.u.), the T' phase persists kinetically.

In summary, we establish that the kinetic persistence of the T' phase
stems from S--S repulsive interactions that impede both nucleation and
interface propagation, and that sulfur vacancies---commonly assumed to
accelerate structural transitions---fail to assist the transformation at
the dominant interface due to their interfacial instability. This
underscores that defect efficacy in phase transformations is governed
not by global concentration, but by local compatibility with the
advancing front, offering a new design criterion for controlling
structural evolution in two-dimensional materials.

\textbf{Methods}

\emph{\textbf{DFT calculations.}}

All the density functional theory (DFT) simulations were implemented in
the Vienna \emph{ab initio} package (VASP).\textsuperscript{31,32} The
generalized gradient approximation (GGA) to exchange-correlation
functional parameterized by Perdew, Burke, and Ernzerhof
(PBE)\textsuperscript{33} was adopted. The projector augmented wave
method\textsuperscript{34} was used to describe core-valence
interaction. A kinetic energy cutoff of 400 eV was used for the
plane-wave basis set. The energy convergence threshold was set to
1×10\textsuperscript{-5} eV. The surface model was separated from its
neighboring images by a vacuum layer larger than 15 Å. For structure
optimization, the conjugate gradient method was used until the force on
each atom was less than 0.01 eV/\AA. The Monkhorst-Pack
method\textsuperscript{35} was used for K-point sampling with spacing
less than 0.025$\times 2 \pi$ \AA \textsuperscript{-1}.

\emph{\textbf{Machine learning potential training.}}

We used DeepMD-kit\textsuperscript{23} to train our MLFF with a cutoff
radius of 6.5 Å for neighbor searching and a fitting net of
\(240 \times 240 \times 240\) for atomic energy evaluation. To
adequately describe the MoS\textsubscript{2} system, we constructed
numerous structures to explore the configuration space. Except for a few
\emph{ab initio} molecular dynamic (AIMD) simulations and structure
optimizations, an iterative scheme was adopted to generate most of the
training set.\textsuperscript{36} That is, we repeatedly train coarse
MLFFs for generating new structures, and label new structures via DFT
calculations to enlarge the current training set for training a better
MLFF in the next round. We have carefully included, in our training set,
possible effects of uniaxial/biaxial stresses, different defects,
different phases, and temperature effects. In addition, we adopted
structures from literature\textsuperscript{37} and our previous research
projects, and used global search methods to find more edge-reconstructed
structures. Our training-set can be classified into the following
categories: \emph{Primitive cell, Supercell, Triangle, Tube, Vacancy,
Edge, Grain boundary (GB), T nucleation, Multilayer, S-deficient phases,
and Other-phases}. A detailed description of these categories is given
in \textbf{Table S1}. We have a total of 37611 structures, containing
2,344,628 force components. We randomly sampled 1000 structures as our
test set, and all others are used to train the final MLFF.

The richness and diversity of the training set are visualized via
principal component analysis (PCA)\textsuperscript{38} of local atomic
environments. As shown in Fig. 2a--b, we applied the SOAP
descriptor,\textsuperscript{39} implemented via the DScribe
library,\textsuperscript{40} to represent each atom's local coordination
as a feature vector. The two dominant principal components (PCA1 and
PCA2) are plotted, colored by the Mo--S coordination number (CN). For
sulfur atoms (Fig. 2a), CN ranges from 0 to 5, with most structures
clustering around CN = 3, which is consistent with the ideal trigonal
prismatic coordination in H/T phases. Similarly, molybdenum atoms (Fig.
2b) exhibit CN from 1 to 8, with a peak at CN = 6, corresponding to
octahedral coordination. The insets highlight representative local
environments, including under-coordinated atoms at edges, vacancies, and
GBs, demonstrating the wide range of chemical environments captured in
the dataset.

\emph{\textbf{MD simulations.}}

ASE\textsuperscript{41} is used to run MD simulations to generate the
training set with coarse MLFF during the iteration process. Typically,
we extract one structure from every 500 steps to further perform DFT
calculations to iteratively enrich our training set. For all phase
transformation related simulations, Lammps\textsuperscript{42} package
is adopted to perform the MD simulation with Nose-Hoover
thermostat.\textsuperscript{43} For these MD simulations, the timestep
is set to be 1 fs.

\emph{\textbf{Free energy calculations}}

The vibrational free energy of H and T phase is calculated based on the
frozen phonon approximation\textsuperscript{44}

\[F_{vib} = \int_{0}^{\infty}{\left\lbrack \frac{\epsilon}{2} + k_{B}T\ln\left( 1 - e^{- \epsilon/k_{B}T} \right) \right\rbrack\sigma(\epsilon)d\epsilon}\]

where \(\epsilon\) is the phonon energy and \(\sigma\) is the phonon
density of states.

\emph{\textbf{Kink-mediated interface propagation: rate-limiting
mechanisms}}

The propagation of a one-dimensional interface, such as a grain boundary
or edge in two-dimensional materials, is governed by the nucleation and
migration of atomic-scale kinks. Kinks are generated in ± pairs at
discrete sites along the interface of length \emph{L}, with a total
generation rate

\[R_{kg} = 2\frac{L}{a}r_{kg}\]

where \emph{a} is the lattice constant parallel to the interface and
\(r_{kg}\) is the nucleation rate at a single site, which is

\[r_{kg} = v \times exp\left( \frac{- \mathrm{\Delta}E_{kg}}{k_{B}T} \right)\]

with \emph{v} \textasciitilde{}
10\textsuperscript{12}-10\textsuperscript{13} Hz,
\(\mathrm{\Delta}E_{kg}\) the kink formation barrier, \(k_{B}\)is
Boltzmann's constant, and \emph{T} is temperature. Each kink nucleation
event increases the total number of elementary kink units
\emph{N\textsubscript{k}} by two---one positive (+) and one negative
(--)---where the sign denotes the local direction of interface
advancement. Kinks migrate along the interface with rate

\[r_{km} = v \times \exp\left( \frac{- \mathrm{\Delta}E_{km}}{k_{B}T} \right)\]

where \(\mathrm{\Delta}E_{km}\) is the migration barrier. Crucially,
annihilation occurs only when kinks of opposite sign meet; collisions
between like-signed kinks result in coalescence into multi-unit steps
but do not reduce \emph{N\textsubscript{k}}, provided
\emph{N\textsubscript{k}} is defined as the total count of elementary
kink units. Under steady-state conditions and low kink density
(\emph{N\textsubscript{k}} $\ll$ \emph{L/a} ), annihilation is dominated
either by kink migration to the fixed ends of the interface or by rare
encounters between opposite-sign kinks. In the end-dominated limit with
very low kink density, balancing generation and annihilation yields

\[N_{k} \approx (L/a)^{2}\left( r_{kg}/r_{km} \right)\]

and in the collision-dominated regime

\[N_{k} \approx 2(L/a)\sqrt{r_{kg}/r_{km}}\]

In both limiting cases, the steady-state kink number depends
exponentially on the difference between the nucleation and migration
barriers through the ratio
\emph{r\textsubscript{kg}/r\textsubscript{km}}. When the nucleation
barrier is significantly larger than the migration barrier---as is
typical in covalently bonded 2D materials like MoS\textsubscript{2} (
\(\mathrm{\Delta}E_{kg} \gg \mathrm{\Delta}E_{km}\)), this ratio becomes
extremely small even at elevated temperatures. Consequently,
\emph{N\textsubscript{k}} remains tiny compared to the total number of
interface sites, confirming that the system operates deep in the
low-density limit.

The propagation velocity of the interface is determined by how rapidly
kinks translate atomic steps forward. Each migrating kink advances the
interface by one perpendicular lattice spacing \emph{b} each time it
moves past an atomic site. With \emph{N\textsubscript{k}} kinks
distributed over \emph{L/a} sites, the average interface advancement per
unit time is

\[R_{ip} = b\frac{N_{k}r_{km}}{L/a}\]

Substituting the steady-state expression for \emph{N\textsubscript{k}}
leads to

\[R_{ip} = br_{kg}L/a\]

demonstrating that the propagation is limited by kink nucleation rather
than migration. This picture is well justified for zigzag-oriented MoS\textsubscript{2}
interfaces (ZZ-Mo\textbar-), where our calculations yield
\(\mathrm{\Delta}E_{kg}\)=1.08 eV and \(\mathrm{\Delta}E_{km}\)=0.20 eV.
At T=300 K, the average kink number on a 1 $\mu$m interface is only
1.5$\times$10\textsuperscript{-8}, rising to approximately 28 at 800 K, which
is still far below \emph{L/a} \textasciitilde{} 10\textsuperscript{4}.
Thus, under typical experimental conditions, kink density remains low,
kink--kink interactions are negligible, and interface motion is
unequivocally dominated by the kinetics of kink generation.

\textbf{Data availability}

The data that support the findings of this study are available from the
corresponding author upon reasonable request.

\textbf{References}

1. Li, W., Qian, X. \& Li, J. Phase transitions in 2D materials.
\emph{Nat. Rev. Mater.} \textbf{6}, 829--846 (2021).

2. Yu, Y. \emph{et al.} High phase-purity 1T'-MoS2- and
1T'-MoSe2-layered crystals. \emph{Nat. Chem.} \textbf{10}, 638--643
(2018).

3. Voiry, D. \emph{et al.} Conducting MoS2 nanosheets as catalysts for
hydrogen evolution reaction. \emph{Nano Lett.} \textbf{13}, 6222--6227
(2013).

4. Liu, L. \emph{et al.} Phase-selective synthesis of 1T' MoS2
monolayers and heterophase bilayers. \emph{Nat. Mater.} \textbf{17},
1108--1114 (2018).

5. Shang, C. \emph{et al.} Superconductivity in the metastable 1 T' and
1 T''' phases of MoS2 crystals. \emph{Phys. Rev. B} \textbf{98}, 184513
(2018).

6. Mak, K. F., McGill, K. L., Park, J. \& McEuen, P. L. The valley Hall
effect in MoS2 transistors. \emph{Science} \textbf{344}, 1489--1492
(2014).

7. Cheng, P., Sun, K. \& Hu, Y. H. Memristive behavior and ideal
memristor of 1T phase MoS2 nanosheets. \emph{Nano Lett.} \textbf{16},
572--576 (2016).

8. Acerce, M., Voiry, D. \& Chhowalla, M. Metallic 1T phase MoS2
nanosheets as supercapacitor electrode materials. \emph{Nat.
Nanotechnol.} \textbf{10}, 313--318 (2015).

9. Duerloo, K.-A. N., Li, Y. \& Reed, E. J. Structural phase transitions
in two-dimensional Mo- and W-dichalcogenide monolayers. \emph{Nat.
Commun.} \textbf{5}, 4214 (2014).

10. Enyashin, A. N. \emph{et al.} New route for stabilization of 1T-WS2
and MoS2 phases. \emph{J. Phys. Chem. C} \textbf{115}, 24586--24591
(2011).

11. Ji, X. \emph{et al.} Interlayer Coupling Dependent Discrete H → T'
Phase Transition in Lithium Intercalated Bilayer Molybdenum Disulfide.
\emph{ACS Nano} \textbf{15}, 15039--15046 (2021).

12. Lin, Y.-C., Dumcenco, D. O., Huang, Y.-S. \& Suenaga, K. Atomic
mechanism of the semiconducting-to-metallic phase transition in
single-layered MoS2. \emph{Nat. Nanotechnol.} \textbf{9}, 391--396
(2014).

13. Zhu, J. \emph{et al.} Argon plasma induced phase transition in
monolayer MoS2. \emph{J. Am. Chem. Soc.} \textbf{139}, 10216--10219
(2017).

14. Nam, D.-H. \emph{et al.} Anion extraction-induced polymorph control
of transition metal dichalcogenides. \emph{Nano Lett.} \textbf{19},
8644--8652 (2019).

15. Li, Z. 1T'-transition metal dichalcogenide monolayers stabilized on
4H-Au nanowires for ultrasensitive SERS detection. \emph{Nat. Mater.}

16. Zhuang, H. L., Johannes, M. D., Singh, A. K. \& Hennig, R. G.
Doping-controlled phase transitions in single-layer MoS2. \emph{Phys.
Rev. B} \textbf{96}, 165305 (2017).

17. Geng, X. \emph{et al.} Pure and stable metallic phase molybdenum
disulfide nanosheets for hydrogen evolution reaction. \emph{Nat.
Commun.} \textbf{7}, 10672 (2016).

18. Kan, M. \emph{et al.} Structures and phase transition of a MoS2
monolayer. \emph{J. Phys. Chem. C} \textbf{118}, 1515--1522 (2014).

19. Zhao, W. \& Ding, F. Energetics and kinetics of phase transition
between a 2H and a 1T MoS2 monolayer---a theoretical study.
\emph{Nanoscale} \textbf{9}, 2301--2309 (2017).

20. Zou, X., Zhang, Z., Chen, X. \& Yakobson, B. I. Structure and
dynamics of the electronic heterointerfaces in MoS2 by first-principles
simulations. \emph{J. Phys. Chem. Lett.} \textbf{11}, 1644--1649 (2020).

21. Jin, Q., Liu, N., Chen, B. \& Mei, D. Mechanisms of semiconducting
2H to metallic 1T phase transition in two-dimensional MoS2 nanosheets.
\emph{J. Phys. Chem. C} \textbf{122}, 28215--28224 (2018).

22. Deringer, V. L., Tchougréeff, A. L. \& Dronskowski, R. Crystal
Orbital Hamilton Population (COHP) Analysis As Projected from Plane-Wave
Basis Sets. \emph{J. Phys. Chem. A} \textbf{115}, 5461--5466 (2011).

23. Zeng, J. \emph{et al.} DeePMD-kit v2: A software package for deep
potential models. \emph{J. Chem. Phys.} \textbf{159}, 054801 (2023).

24. Xiao, Y., Zhou, M., Liu, J., Xu, J. \& Fu, L. Phase engineering of
two-dimensional transition metal dichalcogenides. \emph{Sci. China
Mater.} \textbf{62}, 759--775 (2019).

25. Bourgeois, L., Zhang, Y., Zhang, Z., Chen, Y. \& Medhekar, N. V.
Transforming solid-state precipitates via excess vacancies. \emph{Nat.
Commun.} \textbf{11}, 1248 (2020).

26. Zhang, X. \emph{et al.} Defect-characterized phase transition
kinetics. \emph{Appl. Phys. Rev.} \textbf{9}, (2022).

27. Sekerka, R. F. Equilibrium and growth shapes of crystals: how do
they differ and why should we care? \emph{Cryst. Res. Technol.}
\textbf{40}, 291--306 (2005).

28. Marks, L. D. \& Peng, L. Nanoparticle shape, thermodynamics and
kinetics. \emph{J. Phys. Condens. Matter} \textbf{28}, 53001 (2016).

29. Tang, Q. Tuning the phase stability of Mo-based TMD monolayers
through coupled vacancy defects and lattice strain. \emph{J. Mater.
Chem. C} \textbf{6}, 9561--9568 (2018).

30. Eda, G. \emph{et al.} Coherent atomic and electronic
heterostructures of single-layer MoS2. \emph{ACS Nano} \textbf{6},
7311--7317 (2012).

31. Kresse, G. \& Furthmüller, J. Efficiency of ab-initio total energy
calculations for metals and semiconductors using a plane-wave basis set.
\emph{Comput. Mater. Sci.} \textbf{6}, 15--50 (1996).

32. Kresse, G. \& Furthmüller, J. Efficient iterative schemes for
\emph{ab initio} total-energy calculations using a plane-wave basis set.
\emph{Phys. Rev. B} \textbf{54}, 11169--11186 (1996).

33. Perdew, J. P., Burke, K. \& Ernzerhof, M. Generalized Gradient
Approximation Made Simple. \emph{Phys. Rev. Lett.} \textbf{77},
3865--3868 (1996).

34. Blöchl, P. E. Projector augmented-wave method. \emph{Phys. Rev. B}
\textbf{50}, 17953--17979 (1994).

35. Monkhorst, H. J. \& Pack, J. D. Special points for Brillouin-zone
integrations. \emph{Phys. Rev. B} \textbf{13}, 5188--5192 (1976).

36. Li, P., Zeng, X. \& Li, Z. Understanding High-Temperature Chemical
Reactions on Metal Surfaces: A Case Study on Equilibrium Concentration
and Diffusivity of CxHy on a Cu(111) Surface. \emph{JACS Au} \textbf{2},
443--452 (2022).

37. Zou, X., Zhang, Z., Chen, X. \& Yakobson, B. I. Structure and
Dynamics of the Electronic Heterointerfaces in MoS2 by First-Principles
Simulations. \emph{J. Phys. Chem. Lett.} \textbf{11}, 1644--1649 (2020).

38. Abdi, H. \& Williams, L. J. Principal component analysis.
\emph{WIREs Comput. Stat.} \textbf{2}, 433--459 (2010).

39. Bartók, A. P., Kondor, R. \& Csányi, G. On representing chemical
environments. \emph{Phys. Rev. B} \textbf{87}, 184115 (2013).

40. Himanen, L. \emph{et al.} DScribe: Library of descriptors for
machine learning in materials science. \emph{Comput. Phys. Commun.}
\textbf{247}, 106949 (2020).

41. Hjorth Larsen, A. \emph{et al.} The atomic simulation
environment---a Python library for working with atoms. \emph{J. Phys.
Condens. Matter} \textbf{29}, 273002 (2017).

42. Plimpton, S. Fast Parallel Algorithms for Short-Range Molecular
Dynamics. \emph{J. Comput. Phys.} \textbf{117}, 1--19 (1995).

43. Hoover, W. G. Canonical dynamics: Equilibrium phase-space
distributions. \emph{Phys. Rev. A} \textbf{31}, 1695--1697 (1985).

44. Korotaev, P., Belov, M. \& Yanilkin, A. Reproducibility of
vibrational free energy by different methods. \emph{Comput. Mater. Sci.}
\textbf{150}, 47--53 (2018).

\textbf{Acknowledgements}

We acknowledge funding supports from the Strategic Priority Research
Program of the Chinese Academy of Sciences (Grant Nos. XDB0670000), the
National Natural Science Foundation of China (Grant No. 22403104,
52525201, and 52350209), Science and Technology Commission of Shanghai
Municipality (Grant No. 24CL2900200, 25CL2900200), CAS Project for Young
Scientists in Basic Research (Grant No. YSBR-081).

\textbf{Author Contributions Statement}

P.L. performed DFT, MLFF calculations and MD simulations. Z.D. and F.D.
contributed to the data analysis and supervised the project. All authors
discussed the results and contributed to the manuscript.

\textbf{Competing interests statement}

The authors declare no conflict of interest.

\end{document}